\documentclass[
  twocolumn,
  superscriptaddress,
  amsmath,amssymb,
  aps,
  pra,
  footinbib
]{revtex4-2}
\usepackage{mathptmx}
\usepackage{graphicx}
\usepackage[exfig.port]{adjustbox}

\usepackage{amsmath,amsfonts,bm}

\def\eqref#1{equation~\ref{#1}}

\def\1{\bm{1}}

\DeclareMathAlphabet{\mathsfit}{\encodingdefault}{\sfdefault}{m}{sl}
\SetMathAlphabet{\mathsfit}{bold}{\encodingdefault}{\sfdefault}{bx}{n}

\usepackage{comment}
\usepackage{float}
\usepackage{hyperref}
\usepackage{url}
\usepackage{xcolor}
\newcommand{\bs}{\boldsymbol{\sigma}}
\newcommand{\s}{\sigma}

\begin{document}

\title{Pseudo-likelihood produces associative memories able to generalize,\\even for asymmetric couplings}

\author{Francesco D’Amico}
  \affiliation{Physics Department, Sapienza University of Rome, Piazzale Aldo Moro 5, 00185 Rome, Italy}
  \affiliation{Institute of Nanotechnology, National Research Council of Italy, CNR-NANOTEC, Rome Unit}

\author{Dario Bocchi}
  \affiliation{Physics Department, Sapienza University of Rome, Piazzale Aldo Moro 5, 00185 Rome, Italy}
  \affiliation{Institute of Nanotechnology, National Research Council of Italy, CNR-NANOTEC, Rome Unit}

\author{Luca Maria Del Bono}
  \affiliation{Physics Department, Sapienza University of Rome, Piazzale Aldo Moro 5, 00185 Rome, Italy}
  \affiliation{Institute of Nanotechnology, National Research Council of Italy, CNR-NANOTEC, Rome Unit}

\author{Saverio Rossi}
  \affiliation{Physics Department, Sapienza University of Rome, Piazzale Aldo Moro 5, 00185 Rome, Italy}

\author{Matteo Negri}
  \affiliation{Physics Department, Sapienza University of Rome, Piazzale Aldo Moro 5, 00185 Rome, Italy}
  \affiliation{Institute of Nanotechnology, National Research Council of Italy, CNR-NANOTEC, Rome Unit}

\begin{abstract}
Energy‐based probabilistic models learned by maximizing the likelihood of the data are limited by the intractability of the partition function. A widely used workaround is to maximize the \emph{pseudo‐likelihood}, which replaces the global normalization with tractable local normalizations.  Here we show that, in the zero‐temperature limit, a network trained to maximize pseudo‐likelihood naturally implements an associative memory: if the training set is small, patterns become fixed‐point attractors whose basins of attraction exceed those of any classical Hopfield rule.
We explain quantitatively this effect on uncorrelated random patterns. 
Moreover,  we show that, for different structured datasets coming from computer science (random feature model, MNIST), physics (spin glasses) and biology (proteins), as the number of training examples increases the learned network goes beyond memorization, developing meaningful attractors with non-trivial correlations with test examples, thus showing the ability to generalize.  Our results therefore reveal pseudo‐likelihood works both as an efficient inference tool and as a principled mechanism for memory and generalization.
\end{abstract}

\maketitle

\section{Introduction}

Probabilistic modeling aims to infer the probability distribution of a dataset, in order to both extract relevant features and to generate new, previously unseen samples. 
In this task, it is convenient to parametrize the probability distribution as the exponential of minus an energy function: in this way, the most probable states can be related to the stationary points of the free energy using statistical mechanics, and sampling can be seen as a stochastic process at thermodynamic equilibrium. In machine learning, inferring the parameters of the chosen energy is often referred to as \textit{Boltzmann learning} \cite{ackley1985learning}, or \textit{energy-based probabilistic modeling} \cite{lecun2006tutorial,song2021train}.
In statistical physics, this inference is called an \textit{inverse problem} \cite{nguyen_inverse_2017, inverseProblems}, since the direct problem would be to find the most probable configurations given the couplings of the energy function.

These models can suffer from overfitting, which however can be  difficult to define and quantify \cite{catania2025theoretical}: the many methods available relate to the quality of the generated samples and the moments of their distribution \cite{decelle2025inferring}. One of those methods consist in measuring the propensity to ``memorize" training examples, that is to generate samples from the training set or close to them \cite{bereux2025fast, bonnaire2025diffusion}.

In this work, we adopt this memorization perspective on overfitting and we connect it to the setting of Associative Memories (AMs), also known as Hopfield Networks \cite{amari1977dynamics, hopfield1982neural}. 
AMs are models whose task is to store a set of given examples and, when presented with an input that is a noisy version of one example, to output the corresponding denoised version. This task is often achieved by building an energy function whose minima (the ``memories") are close to the desired examples, so that when the model is initialized within the basin of attraction of a specific minimum the dynamics of the system will naturally converge to the memory. 

AMs have been traditionally regarded as conceptually important but impractical, due to their limitations of small capacity \cite{amit1987statistical} and difficulty in dealing correlated examples \cite{van1997hebbian}. In the recent years, both difficulties have been overcome. 
First, the so-called dense AMs \cite{krotov2016dense} have been shown to have polynomial or even exponential capacity in the size of the system \cite{lucibello2023exponential}, and found applications in different contexts of modern machine learning, such as the attention mechanism \cite{ramsauer2020hopfield}, generative diffusion \cite{ambrogioni2024search} and probabilistic modeling \cite{schaeffer2024bridging}.
Second, it was shown that even the simplest AMs can exploit correlations in the data to enter a generalization phase, where they produce minima of the energy close to previously unseen configurations that have the same distribution as the training examples \cite{negri2023storage,kalaj2024random,serricchio2025daydreaming}. 

The newfound capability of AMs to generalize is the main motivation for this work: while the interpretation of overfitting in probabilistic models as memorization is intuitive, the connection with AMs may also provide a novel perspective on generalization. 

The classic inference of energy-based probabilistic models by maximizing the likelihood of the data \cite{fisher1997absolute,mackay2003information} is limited by the difficulty of estimating the partition function. 
A common strategy to avoid this problem is to maximize the \textit{pseudo-likelihood} instead \cite{besag1974spatial,nguyen_inverse_2017, inverseProblems}, that only requires tractable normalizations. These method has seen widespread application, for instance, in the inference of protein sequences \cite{Ekeberg2013,Ekeberg2014}.

In this work we study energy-based models trained by maximizing the pseudo-likelihood and their dynamics in the limit of zero temperature.
The contributions of this work can be summarized as follows.
\begin{itemize}
    \item We show that, for small enough training sets, maximizing the pseudo-likelihood produces an AM, consolidating the intuitive perspective of overfitting as memorization of the training set. A noteworthy results is that the same holds for asymmetric couplings. We explain this fact using the theory of spherical perceptrons.
    \item We show that by increasing the size of the training set the AM produced by pseudo-likelihood enters a ``generalization" regime, where training and test examples are at the same distance from fixed-points of the AM. We use this regime to propose a new perspective to quantify generalization in energy-based probabilistic models: we measure the correlation between the fixed-points of the AM and the test examples, knowing that if the dataset is simple enough (or the model is powerful enough) this correlation may reach one.
    \item We support this perspective with numerical results using the simplest possible architecture, to understand fundamental behaviors of memorization and generalization that may also appear with more complicated architectures. We present results on a variety of datasets from computer science (random feature model, MNIST), physics (spin glasses) and biology (proteins).
\end{itemize}

This paper is organized as follows. In section~\ref{sec:PL} we review the inference via pseudo-likelihood maximization and we describe our model. In section~\ref{sec:PLisAM} we describe the quantitative connection between pseudo-likelihood maximization and AMs. In section~\ref{sec:results} we detail the numerical results on the various dataset that we considered. In section~\ref{sec:conclusions} we discuss the significance of our work and we propose future perspectives.

\section{General setting of Pseudo-Likelihood}
\label{sec:PL}

We first revise the maximum likelihood principle to fix the notation and to introduce the pseudo-likelihood method. In the next section we show why this model has a natural interpretation as an associative memory.

\paragraph*{Maximum-likelihood principle}
Let $\mathcal{D}=\{\bm\xi^{\mu}\}_{\mu=1}^{P}$ be a dataset of examples $\bm\xi=\{\xi_{i}\}_{i=1}^{N}$ that we want to model with a probability distribution $p_J(\bm{x})=\exp\{-\lambda E_J(\bm{x})\}/Z_J$ over a set of variables $\bm{x}=\{x_{i}\}_{i=1}^{N}$ dependent on a set of parameters $J$, where $Z_J=\int d\bm{y}\, \exp\{-\lambda E(\bm{y})\}$ is the partition function,  $E_J(\bm{x})$ is the energy function and $\lambda$ is the inverse temperature. The \textit{likelihood} of a data point is $p_J(\bm\xi^{\mu})$. One can infer $J$ by maximizing the likelihood of all the data points simultaneously \cite{fisher1997absolute,mackay2003information}, which results in the minimization of the negative log-likelihood (NLL) loss
$
    \mathcal{L}=-\sum_{\mu=1}^{P}\log p_J(\bm\xi^{\mu}).
$
This loss is difficult to optimize, because it requires the estimation of the normalization $Z_J$. One of the most common strategies to deal with this term is the family of algorithms related to the minimization of Contrastive Divergence \cite{hinton_training_2002}, which exploit fast out-of-equilibrium stochastic processes instead of a standard Monte Carlo processes, which are usually inefficient as they reach equilibrium very slowly \cite{agoritsas_explaining_2023} 
(recent advances \cite{bereux2025fast} allow to reach the equilibrium process much faster in Restricted Boltzmann Machines).

\paragraph*{Pseudo-likelihood}
Another strategy to avoid the estimation of $Z_J$ entirely is to 
approximate the joint probability
as the product of conditionals $p_J(\bm{x})\simeq\prod_{i}p_{i}(x_{i}|\bm{x}_{\setminus i})$, where $\bm{x}_{\setminus i}=\{x_{j}\}_{j(\neq i)}$ is the set of all variables except the $i$-th variable. 
The advantage is that now the conditionals can be written as\footnote{To lighten the notation, we omit the dependence on $J$ of $\psi(\bm{x})$, $p_{i}(x_{i}|\bm{x}_{\setminus i})$ and $Z_{i}(\bm{x}_{\setminus i})$.}
$p_{i}(x_{i}|\bm{x}_{\setminus i})=\exp\{-\lambda E_J(\bm{x})\}/Z_i(\bm{x}_{\setminus i})$, where the normalization $Z_i(\bm{x}_{\setminus i})=\int d\bm{y}_i\,\exp\{-\lambda E({y}_i,\bm{x}_{\setminus i})\}$ requires only a single integral and therefore is tractable.
The likelihood of a data point within this ansatz is called \textit{pseudo-likelihood} \cite{besag1974spatial}.
If we plug this expression in the NLL loss we get a \textit{negative log-pseudo-likelihood} (NLpL) loss
\begin{equation}
  \mathcal{L}=-\sum_{\mu=1}^{P}\sum_{i=1}^N\log p_{i}(\xi_{i}^{\mu}|\bm{\xi^{\mu}}_{\setminus i}),  
  \label{eq:loss}
\end{equation}
which is the quantity that we minimize to train the model in all the experiments described in this work.

\paragraph*{Gibbs sampling and memory retrieval}
The core idea of a probabilistic model is the possibility of sampling from the inferred distribution.
The standard sampling used in the context of pseudo-likelihood is Gibbs sampling \cite{geman1984stochastic}, which consists in updating the variables using the conditional probabilities: in practice, at time $t$ one picks a variable $x_i$ to update, then fixes the values of all the other variables, and then samples
\begin{equation}
    x_i^{(t+1)} \sim  p_{i}(x_{i}|\bm{x}_{\setminus i}^{(t)}).
    \label{eq:dynamics}
\end{equation}

This setting is also known as a stochastic neural network and it has been connected to an online optimization of pseudo-likelihood in \cite{saglietti_statistical_2018}. 
Rather than using Eq.~\ref{eq:dynamics} for sampling, in this work we will study the recurrent update as an AM and consider its storage and retrieval capabilities. To simplify the analysis of AMs, we study the update in Eq.~\ref{eq:dynamics} at zero temperature ($\lambda\to\infty$), so that we can check that we arrived at a fixed point and avoid sampling.
In the limit $\lambda\to\infty$ the update in Eq.~\ref{eq:dynamics} becomes: 
\begin{equation}
    x_i^{(t+1)} = \text{argmax}_{x_i} \, p_i(x_{i}|\bm{x}_{\setminus i}^{(t)}).
    \label{eq:dyn_p_zero_T}
\end{equation}
Given the set of configurations $\{\bm{\xi^\mu}\}$ with $\mu=1,...,P$, we say that the system \textit{stores} them if they are fixed points of Eq.~\ref{eq:dyn_p_zero_T}. We also say that the model \textit{retrieves} a memory if it converges to its starting from its corrupted version. We refer to the ratio $\alpha = P/N$ as the \textit{model load}.
In this perspective it is natural to study overfitting by asking if training examples are fixed points. In the same way, we will also study generalization by asking if test examples are fixed points.
At variance with Gibbs sampling, we run the dynamics with parallel updates (all variables at the same time) instead of sequential updates (one variable at the time). We make this choice because  parallel updates is significantly faster, and because we did not notice any difference between the two modes in the memory retrieval. 
Note that, if the dynamics is parallel instead of sequential (and if $J$ is symmetric), the model samples from a distribution that is different from $p_J(\bm{x})=\exp\{-\lambda E_J(\bm{x})\}/Z_J$ (described in \cite{peretto_collective_1984}), but since here we are only interested in memory retrieval we can ignore this difference.

\paragraph*{Energy-based pseudo-likelihood} As we will see in the next section, the energy-based parametrization of pseudo-likelihood provides a natural setting to understand the AM properties of the recurrent dynamics in Eq.~\ref{eq:dyn_p_zero_T}. 
In this work we consider a model with two-body interaction $E(\bm x)=-\sum_{i\neq j}J_{ij}x_i x_j$ and binary variables $x_i\in\{\pm1\}$. With this choice, we can rewrite the NLpL loss in Eq.~\ref{eq:loss} as
\begin{equation}
\mathcal{L}=-\sum_{i,\mu}\Bigl[\lambda\,\xi_i^\mu\!\sum_{j\neq i}\!J_{ij}\xi_j^\mu-\log 2\cosh\!\bigl(\lambda\!\sum_{j\neq i}\!J_{ij}\xi_j^\mu\bigr)\Bigr] \label{eq:loss_ising}
\end{equation}
and the update in Eq.~\ref{eq:dyn_p_zero_T} as
\begin{equation}
    x_i^{(t+1)} = \mathrm{sign}(\sum_{j (\neq i)} J_{ij} x_j^{(t)}),
    \label{eq:dyn_E_zero_T}
\end{equation}
which corresponds to the zero-temperature limit $\lambda\to\infty$ of the dynamics in Eq.~\ref{eq:dynamics}. Note that Eq.~\ref{eq:dyn_E_zero_T} also reads as the update of a recurrent network built with perceptrons, where each perceptron is parametrized by a row $\bm{J}_i$ of the coupling matrix $J$.
With this perspective, saying that we are interested in training examples as fixed points corresponds to saying that each perceptron should correctly classify the training examples, i.e. we want to solve the self-supervised problem of finding $\bm{J}_i$ such that 
\begin{equation}
    \xi_i^{\mu} = \mathrm{sign}(\sum_{j (\neq i)} J_{ij} \xi_j^{\mu})
    \label{eq:perceptrons}
\end{equation}
for all $i,\mu$, which corresponds to the training of $N$ independent perceptrons. This is also reflected by the fact that the NLpL loss in Eq.~\ref{eq:loss_ising} is factorized with respect to index $i$, and can be written as $\mathcal{L}=\sum_i\ell_i$. These observations will be the core of section~\ref{sec:PLisAM}.

\paragraph*{The training procedure}
We train the couplings matrix $J$ by minimizing loss in Eq.~\ref{eq:loss_ising} with gradient descent. We stop the training when the size of the basins of attraction of the training examples do not change anymore (the gradient keeps increasing the norm of $J$ even after the size of the basins converge). A factorized loss allows each row of $J$ to be trained in parallel, giving a substantial numerical advantage.
The couplings inferred with this method are in general not symmetric, and therefore they do not produce an energy function when inserted back in $E_J(\bm{x})$. Since energy is a desirable concept, strategies to symmetrize the couplings are commonly used in this kind of inference \cite{Ekeberg2013,Ekeberg2014,nguyen_inverse_2017}. In the following, instead, we keep the couplings asymmetric, and we will see that they still produce AMs.
With this choice, instead of a minimizing a global energy function, each variable $x_i$ independently aligns to the corresponding  local field $h_i(\bm{x}_{\setminus{i}})=\sum_{j(\neq k)}J_{kj}\xi_{j}^{\mu}$, which may be interpreted as the optimization of a \textit{local} energy term.

\begin{figure*}
    \centering
    \includegraphics[width=0.49\linewidth]{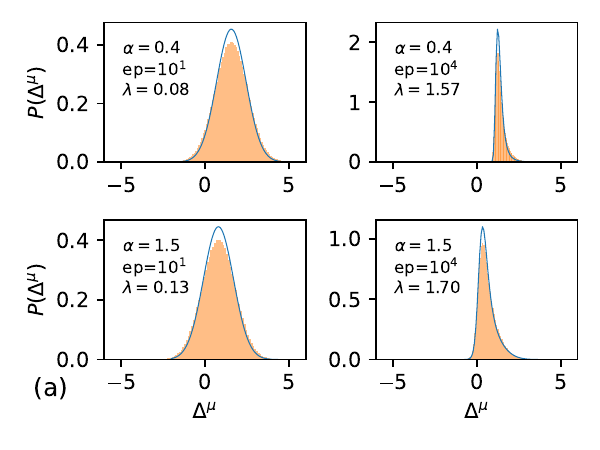}
    \includegraphics[width=0.49\linewidth]{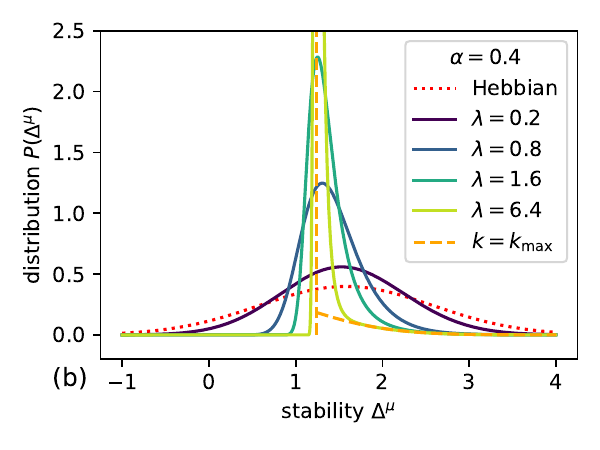}

    \caption{\textbf{Maximizing pseudo-likelihood goes from Hebbian learning to perceptrons with maximum margins.} \textit{Panel (a)}: Comparison between the analytical distribution at fixed $\lambda$ (blue lines) and the numerical distribution with free norm at $\lambda_\mathrm{free}(t ) = \lambda$ (orange histogram), for different loads $\alpha$ and training epochs. While the two distributions come from different settings and do not exactly match, key properties for retrieval such as the mean, minimum, and variance vary similarly as $\lambda$ increases.
    \textit{Panel (b)}: The distribution of margins in the perceptron spherical perceptron trained maximizing pseudo-likelihood, as a function of the parameter $\lambda$. The load is fixed to $\alpha=0.4$.
    }
    \label{fig:P_delta}
\end{figure*}

\section{Why Pseudo-Likelihood produces Associative memories}
\label{sec:PLisAM}

In this section we expand the observation around Eq.~\ref{eq:perceptrons}, using properties of independent perceptrons to describe the connections between the maximization of pseudo-likelihood and the production of AMs. We first sketch the discussion, then we provide technical details.
\begin{itemize}
    \item The first step is to recognize that the loss function of each perceptron can be written as a logistic loss \cite{besag1974spatial}:
    \begin{equation}
        \ell_i(\bm{J}_i) = \sum_{\mu} \log\left(1 + e^{ -\lambda \xi_i^\mu \sum_{j \ne i} J_{ij} \xi_j^\mu } \right)
        \label{eq:logistic_loss}
    \end{equation}
    which is known to have an implicit bias \cite{Soudry2018,Montanari2024NegativePerceptron}: the training converges to a weight vector $\bm{J}_i$ that maximizes the classification margins $k$ (compatibly with the load $\alpha$ used, see \cite{gardner1988space}). This means that maximizing pseudo-likelihood, for small enough $\alpha$, solves the problem of finding $\bm{J}_i$ such that 
\begin{equation}
    \xi_i^{\mu} = \mathrm{sign}(\sum_{j (\neq i)} J_{ij} \xi_j^{\mu} - k)
    \label{eq:perceptrons_with_margins}
\end{equation}
for all $i,\mu$ and for the maximum possible $k$.
    \item The second step is to note that training self-supervised perceptrons with large classification margins indeed produces an AM \cite{forrest1988content, benedetti2022supervised}. 
    \item Finally, we approximate the couplings $J$ at finite training times with solutions of the problem at the corresponding fixed norm \cite{aubin2020generalization}, as it was shown that the \textit{whole} training process has an implicit bias towards solutions at fixed norm \cite{d2025implicit}. This allows us to show that the model implements Hebbian learning at small training time and increases the storage capacity as the learning continues. The analysis for finite training time is especially important for practical cases, as numerically the convergence towards the maximum margin solution is very slow \cite{Soudry2018,Montanari2024NegativePerceptron}. Moreover, terms that regularize the norm are often added to the loss, which prevent the divergence and act as an effective way to fix the norm (the results described in section \ref{sec:res_proteins} use a regularization term). 
\end{itemize}

These connections are confirmed by the numerical results described in section~\ref{sec:res_uncorr}.

\paragraph*{Technical discussion}
We consider random uncorrelated binary examples $\xi_i^\mu=\pm1$, i.i.d. (indipendent and identically distributed) according to a symmetric Rademacher distribution, i.e. $P(\xi_i = 1)= P(\xi_i = -1) = 0.5$.
Let us define the \textit{stability} $\Delta_i^\mu$ of variable $i$ of the example $\bm{\xi}^\mu$ as 
\begin{equation}
\Delta_i^\mu= \xi_i^\mu\sum_{j\neq i}J_{ij}\xi_j^\mu.
\end{equation}
The stability $\Delta_i^\mu$ is directly related to the retrieval of pattern $\bm{\xi}^\mu$: if $\Delta_i^\mu >k$, then the model satisfies Eq.~\ref{eq:perceptrons_with_margins} with margin $k$. If this is true $\forall i$, then $\bm{\xi}^\mu$ is a retrievable memory of the model.

We can rewrite NLpL loss in Eq. \ref{eq:loss_ising} as $\mathcal{L}=\sum_i\ell_i$, where
\begin{equation}
\ell_i=\sum_{\mu}V_\lambda(\Delta_i^\mu),\qquad V_\lambda(\Delta)=\lambda\Delta-\log 2\cosh(\lambda\Delta). \label{eq:single_perceptron_loss}
\end{equation}

Eq. \ref{eq:single_perceptron_loss} allows us to map the minimization directly onto $N$ independent classical perceptron storage problems, where standard statistical-mechanics techniques \cite{gardner1988space, engel2001statistical} allow to obtain the margin distribution $P_\lambda(\Delta)$ for the minimizer of the loss. We report the derivation in Sec.~\ref{sec:replicaGardner} of the Appendix.  We remark that the NLpL loss in Eq. \ref{eq:loss_ising} is minimized independently by each row $\bm{J}_i$ of the coupling matrix, so we consider only a single $i$. The same argument can then be replicated to all $i$ values.Therefore, we study weights vector $\bm{J}\in \mathbb{R}^N$ and stabilities $\Delta^\mu$, dropping the dependence on index $i$.

The derivation requires that we fix the norm of the weights vector $\bm{J}$ to $|\bm{J}|^2 = 1$, while the actual numerical optimization of the loss function in Eq.~\ref{eq:loss_ising} happens with free norm. 
Since $V_\lambda$ depends only on the product $\lambda\Delta$—up to an overall factor that does not affect the location of the minimizers—and that $\Delta \propto \|\bm J\|$, tuning $\lambda$ at fixed norm is equivalent to keeping $\lambda = 1$ and varying the norm. Using this equivalence, we can compare the static analytical predictions to numerical simulations where the perceptron is trained with $V_{\lambda=1}$ without norm constraints, so that the norm plays the role of an effective $\lambda$. 
Indeed, in Fig.~\ref{fig:P_delta}a we observe that the training dynamics changes the stability distribution in a way similar to how the parameter $\lambda$ affects the analytic distribution obtained at fixed norm.

There are two relevant asymptotic behaviors of the $\lambda$ parameter, as showed in Fig.~\ref{fig:P_delta}b:
\begin{enumerate}
    \item  $\lambda \to 0$: the potential reduces to $V_\lambda(\Delta) \simeq -\lambda \Delta$, recovering Hebbian learning; $P_\lambda(\Delta)$ is a Gaussian (Fig.~\ref{fig:P_delta}b, red dotted line). 
    \item $\lambda \to \infty$: solving the model at finite $\lambda$ 
    and subsequently taking $\lambda \to \infty$ yields to a distribution that approaches the maximally stable perceptron, defined by $\bm{J}_{\mathrm{ms}} = \arg\max_{\bm J}\min_\mu \Delta^\mu(\bm J)$; $P_\lambda(\Delta)$ is a Gaussian truncated at $\Delta=k$ (Fig.~\ref{fig:P_delta}b, orange dashed line).
\end{enumerate}
Thus, increasing \(\lambda\) continuously shifts the objective from maximizing the \emph{average} margin ($\ell=\lambda \sum_\mu  \Delta ^\mu$ when $\lambda=0$) to maximizing the lowest margin (via the implicit bias \cite{Soudry2018,Montanari2024NegativePerceptron}). 

Note that, for $\lambda<\infty$, the distribution  $P_\lambda(\Delta)$ always has a negative tail. This tail is responsible for the finite storage capacity of the model, since it implies that a fraction of the variables are misaligned with the desired example. If this tail is not too large, the model is nevertheless able to align part of the misaligned variables with the recurrent iterations, as long as the load is below a critical value $\alpha_\mathrm{crit}$ (i.e. the capacity of the model). For an Hopfield model,  $\alpha_\mathrm{crit}\simeq0.14$, and the attractors have a typical overlap of 0.967 with the examples at the critical point \cite{amit1987statistical}.

From the changes in $P_\lambda(\Delta)$ when we increase $\lambda$ we can infer how the model is changing during training: the average stability decreases, but the negative tail shifts towards higher values. Since it is this tail that determines the capacity $\alpha_\mathrm{crit}$, this picture suggests that at the beginning of the training process the capacity increases, while at later stages the training just optimizes classification margins to have a sharp minimum at the maximum value allowed by the load $\alpha$ at which we train \cite{gardner1988space}.

This analysis assumes random, uncorrelated data, but provides a clean framework for how the pseudolikelihood is able to create an AM. In the next sections we test the effectiveness of pseudo-likelihood-based AMs on structured inputs by evaluating the model's performance on correlated examples, and in particular on real-world datasets. This question is especially important given the recent works show that AMs, when dealing with correlated examples, store them more efficiently than just building attractors close to the training examples  \cite{negri2023storage,kalaj2024random,serricchio2025daydreaming}.

\begin{figure*}
\centering
\includegraphics[width=0.49\textwidth ]{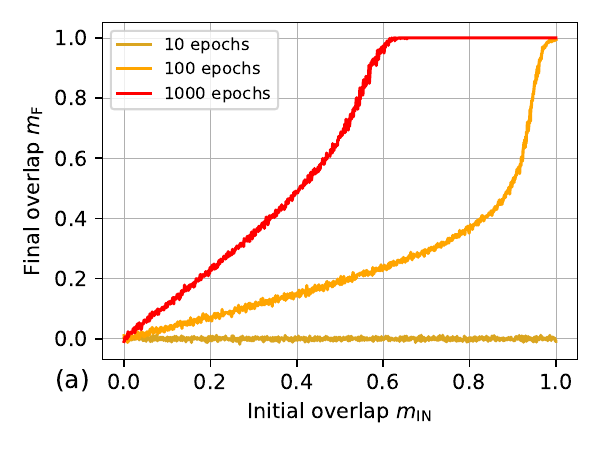}
\includegraphics[width=0.49\textwidth ]{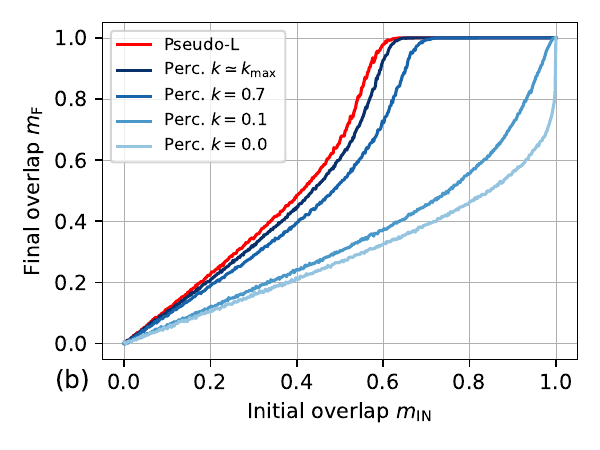}

\caption{\textbf{Maximizing pseudo-likelihood produces basins around uncorrelated training examples.} \textit{Panel (a)}: retrieval map of random i.d.d. examples during the training process for $\alpha=0.4$. The error bars correspond to the average over the whole dataset for a single instance of training. \textit{Panel (b)}: retrieval map of random i.d.d. examples at the end of the training, comparing the maximization of pseudo-likelihood (at 1000 epochs) and perceptrons with different values of the margin.}
\label{fig:uncorr}
\end{figure*}

\section{Numerical results}
\label{sec:results}

We train the model as described in section \ref{sec:PL} on multiple datasets. Then, on each dataset we test the dynamics in Eq. \ref{eq:dyn_E_zero_T} in the following way: we initialize the dynamics starting from a configuration $\bm{x}^\mathrm{IN}$ with overlap $m_\mathrm{IN}=\bm{x}^\mathrm{IN}\cdot \bm{x}^*/N$ with the chosen one, and then we update it until we reach a fixed point $\bm{x}^\mathrm{F}$.  Finally, we measure the overlap $m_\mathrm{F}=\bm{x}^\mathrm{F}\cdot \bm{x}^*/N$. 

The main experiment that we show is the measure of the stability of training, test and random examples at different loads, by starting at $m_\mathrm{IN} = 1$ and measuring $m_\mathrm{F}$ as a function of $\alpha$. We will systematically study basins of attraction (that is, varying also $m_\mathrm{IN}$) only for uncorrelated data.
Our goal is to show two different phases, as highlighted in \cite{kalaj2024random}: a \textit{storage phase}, where only training examples are fixed point of the dynamics, and a \textit{generalization phase}, where attractors appear near previously unseen examples;

We first show results on synthetic datasets where theoretical thresholds are known \cite{amit1987statistical,kalaj2024random} and then move to two real datasets to check if the same phenomenology is present.

We will see that the numerical results for uncorrelated data confirm the picture described in section \ref{sec:PLisAM}, that PL increases the capacity of the AM. This is also true for synthetic correlated data: the learning and generalization phases are extended. When we move to real datasets, we will find a generalization phase there too. On MNIST we are able to asses the quality of the generalization by visually inspecting the attractors. On proteins and spin glasses, this assessment is not possible, but the qualitative behavior is the same.

\begin{figure}
    \centering
    \includegraphics[width=0.49\textwidth ]{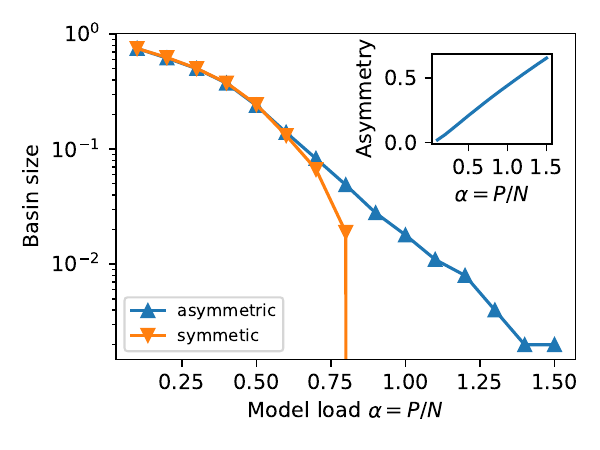}
    \caption{\textbf{Pseudo-likelihood produces large basins of attraction of uncorrelated examples, even with asymmetric couplings.} We compare the size of the basins of attraction of asymmetric and symmetric couplings, for various $\alpha$. The size of the basins is computed as $1-m_\mathrm{IN}^\mathrm{max}$, where $m_\mathrm{IN}^\mathrm{max}$ is the point where the retrieval map goes below $m_\mathrm{F}=0.99$. \textit{Inset}: we plot the asymmetry of the coupling matrix $\Vert J-J^\top\Vert_2/\Vert J\Vert_2$ as function of $\alpha$. In all panels the dimensionality of the data is $N=1000$.} 
    \label{fig:res_uncorr}
\end{figure}

\begin{figure*}
\centering
\includegraphics[width=0.99\textwidth ]{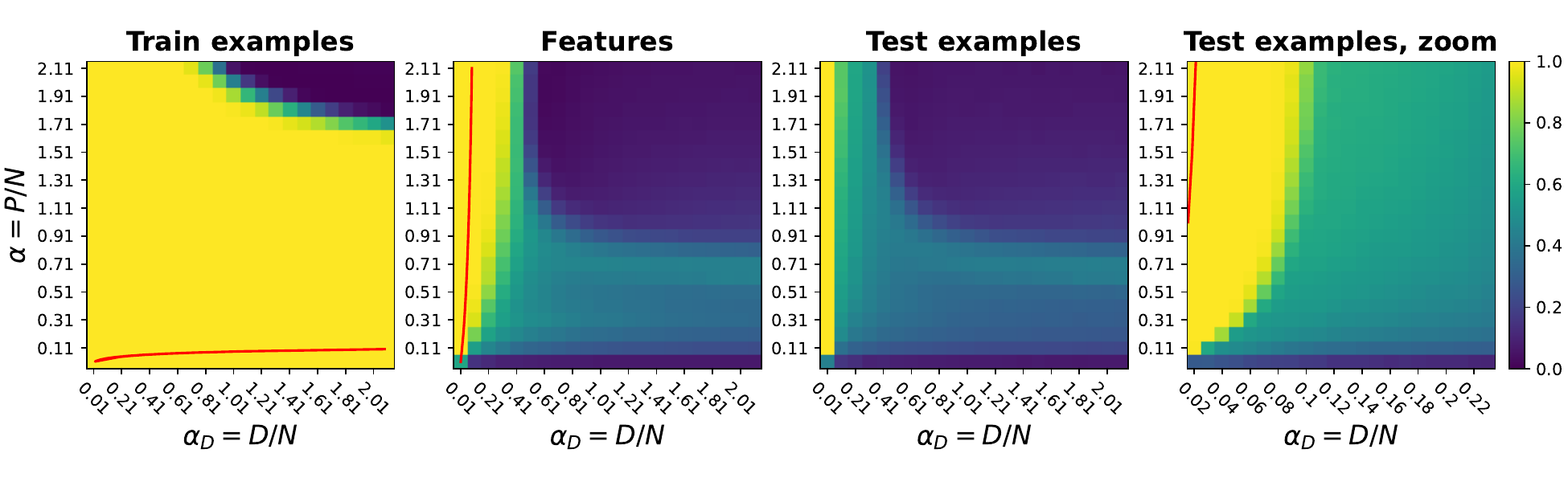}

\caption{\textbf{Pseudo-likelihood improves generalization for correlated random-features examples.} The colors correspond to the values of the final overlap $m_\mathrm{F}$ (computed with respect to train examples, features and test examples) reached starting from $m_\mathrm{IN}=1$, corresponding to storage, learning and generalization phases. Pseudo-likelihood enhances the area of all three phases with respect to Hebbian learning, shown as red curves taken from \cite{kalaj2024random}. The dimensionality of the data is $N=1000$ in the fist three panels and $N=2000$ in the last one. $L=3$ in all panels.}
\label{fig:corr}
\end{figure*}

\subsection{Uncorrelated synthetic examples} 
\label{sec:res_uncorr}
We train the model with random binary i.i.d. (independent and identically distributed) examples, with the components $\xi_i^\mu=\pm1$ distributed according to a symmetric Rademacher distribution, i.e. $P(\xi_i = 1)= P(\xi_i = -1) = 0.5$. 
We test the basin of attraction of training examples. To do so, we plot $m_\mathrm{F}$  as a function of $m_\mathrm{IN}$ for different values of $m_\mathrm{IN}$ (\textit{retrieval map}). 

In Fig.~\ref{fig:uncorr}a we plot retrieval maps during the training process, giving a numerical confirmation that the model becomes an AM.  In Fig.~\ref{fig:uncorr}b we compare the retrieval maps produced by perceptrons trained with various values of $k$: we show that the retrieval maps produced by pseudo-likelihood is close to the retrieval map with $k\simeq k_\mathrm{max}$ (the discrepancy is due to the slow convergence of the perceptron rule near $k_\mathrm{max}$ \cite{gardner1988space, Soudry2018,Montanari2024NegativePerceptron}).

In Fig.~\ref{fig:res_uncorr} we plot the size of the basins of attraction for different values of $\alpha$, computed as $1-m_\mathrm{IN}^\mathrm{max}$, where $m_\mathrm{IN}^\mathrm{max}$ is the point where the retrieval map goes below $m_\mathrm{F}=0.99$.
Our training procedure yields large basins of attraction around training examples well above the capacity $\alpha_c$ of an Hopfield model ($\alpha_c$) \cite{hopfield1982neural} (also compare with the retrieval maps in Fig.~\ref{fig:uncorr} at $\alpha = 0.4$). The size of the basins rapidly approaches zero for $\alpha>1$ (the theoretical bound for asymmetric couplings is $\alpha=2$ \cite{gardner1988space}) as shown in Fig.~\ref{fig:res_uncorr}. Notably, the basins in the asymmetric case are equal or larger than those for symmetric couplings.

\subsection{Correlated synthetic examples}
We train the model with binary examples $\xi_i^\mu$ generated as superposition of binary random features $f_{ki}=\pm1$ i.i.d. with uniform probability, namely $\xi_i^\mu=\mathrm{sgn}(\sum_{k=1}^D c^\mu_k f_{ki} )$. For fixed $\mu$, the coefficients $c_k^\mu \in \{-1,+1\}$ have $L$ non-zero entries in random locations and uniformly sampled in $\pm1$, so that each example is the superposition of $L$ features. This structure of patterns was proposed in \cite{mezard2017mean, Goldt_2020,gerace2020generalisation,Baldassi2022Learning} as a model for the \textit{Hidden Manifold Hypothesis} of real datasets, where high-dimensional inputs $\xi_i^\mu$ that lie on a lower-dimensional manifold of latent variables $c_k^\mu$. We studied the sparse case $L=\mathcal{O}(1)$, in which the Hopfield model is able to store patterns in the $D\propto N$ regime \cite{kalaj2024random}. We can interpret the $L=\mathcal{O}(1)$ condition as the hypothesis that data-points of a real dataset are a combination of a number of features that does not depend on the size of the system.
We see from Fig.~\ref{fig:corr} that the phase diagram of networks trained via pseudo-likelihood shows bigger storage and generalization phases with respect of those of the Hopfield model (see \cite{kalaj2024random}). Note that a region where features, training and test patterns are stable. Notice (from the first panel of Fig.~\ref{fig:corr}) that, due to data being correlated, it is possible to store them above the theoretical threshold $\alpha=2$  for uncorrelated examples \cite{gardner1988space}.

\subsection{MNIST}
\begin{figure*}
\centering
\includegraphics[width=0.49\textwidth]{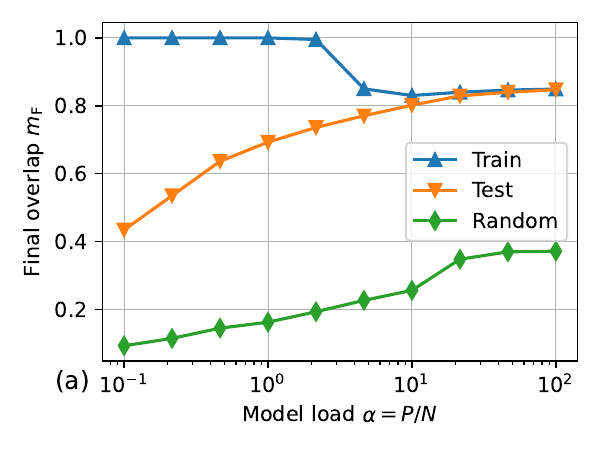}
\raisebox{0.06\height}{\includegraphics[width=0.49\textwidth]{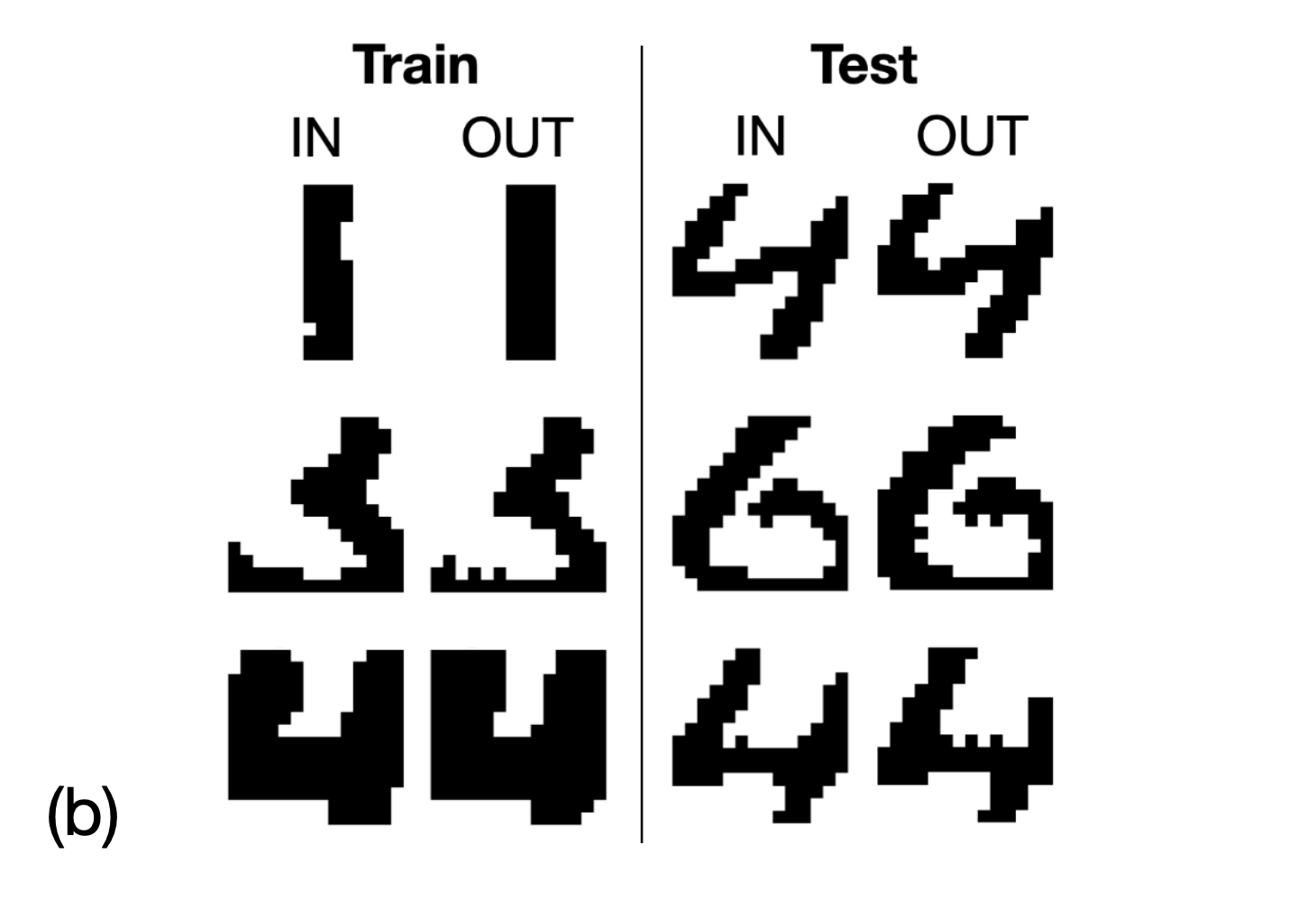}}

\caption{
\textbf{Generalization transition for MNIST data.} \textit{Panel (a)} We show final overlap of dynamics starting from training or test examples, as function of dataset size $\alpha= P/N$. 
\textit{Panel (b}: binarized $14 \times 14$ MNIST dataset. Together with train and test examples, we also show the curve for images that are random sets of pixels (i.i.d configurations on the $N$-dimensional hypercube), to check if the network possess fixed-points with high correlation even for random data, an effect that would suggest that the network is not working properly. Error bars are negligible.
}
\label{fig:mnist}
\end{figure*}

As a simple instance of real-world dataset, we train the model using the binarized $14\times14$ MNIST described in \cite{belyaev2020classification}. In Fig.~\ref{fig:mnist} we show that pseudo-likelihood stores training images for a small value of $\alpha = {P}/{N}$ (storage phase), $P$ being the size of training dataset. For bigger values of $\alpha$ train and test images produce a final overlap $m_{\mathrm{F}}\simeq 0.85$. By inspecting visually retrieved examples, we see that this value of the final overlap corresponds to a very good attractors, i.e. close to the input even for previously unseen examples (see Fig.~\ref{fig:mnist}b and Fig.~\ref{fig:mnsit_examples} in appendix~\ref{sec:appendix_mnist}).

\subsection{Protein sequences}
\label{sec:res_proteins}
\begin{figure*}
    \centering
    \includegraphics[width=0.49\linewidth]{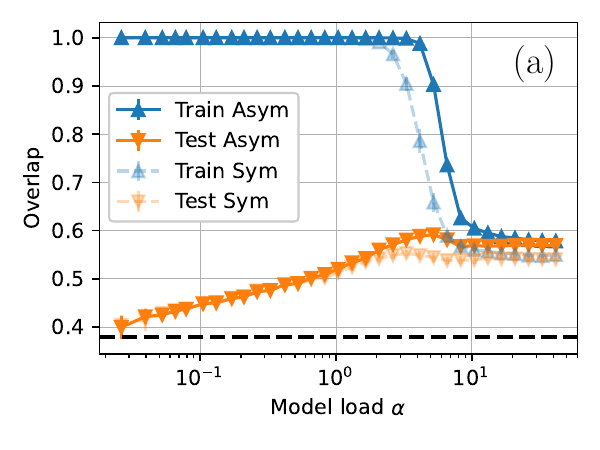}
    \includegraphics[width=0.49\linewidth]{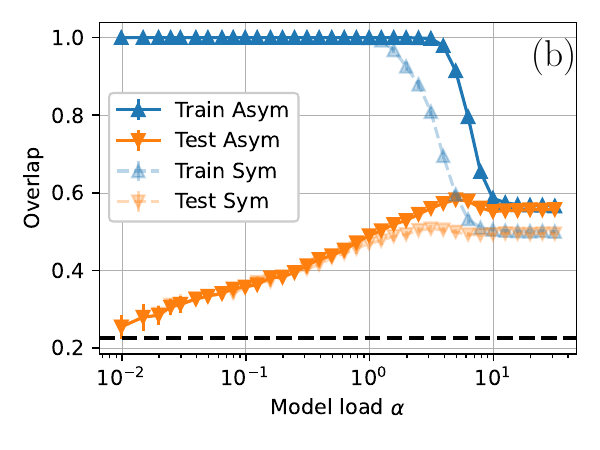}
    \caption{\textbf{Generalization transition for protein sequences.} We show final overlap of dynamics starting from protein sequences from the families DNA-Binding Domain (panel (a)) and the Beta Lactamase (panel (b)). Blue and red correspond to a model with asymmetric couplings, purple and yellow with symmetric couplings. For both, we show train and test examples. The black dashed line is the average overlap $\langle O \rangle$ between natural sequences. Results are averaged over $20$ choices of the training set for each value of $\alpha$.}
    
    \label{fig:Proteins}
\end{figure*}

We consider biological sequences, focusing in particular on proteins. 
Proteins are chains of amino acids and play essential roles in many biological processes. 
They can be grouped into families based on their biological functions; however, substantial sequence variability often exists within each family.

A growing area of interest is the development of generative models capable of producing novel protein sequences that differ from natural ones at the amino acid level, while preserving their biological function. 
A variety of models with diverse architectures have been proposed to generate artificial sequences within a given protein family \cite{Figliuzzi2018,cocco2018inverse,tubiana2019,trinquier2021efficient}, and some have proved successful in producing functional variants \cite{russ2020evolution}. 
Among these approaches is \textit{plmDCA}  \cite{Ekeberg2013,Ekeberg2014}, a model based on pseudo-likelihood maximization  originally developed for contact prediction, which has also been applied to sequence generation.
In \textit{plmDCA} the probability of observing a specific amino acid $a$ at site $i$, given the rest of the sequence $\textbf{a}_{ \setminus i}$, is defined as 
\begin{equation}
    \label{eq:plmDCA}
    P(a_i =a | \textbf{a}_{\setminus i}) \propto \exp \left( h_i(a) + \sum_{j \neq i}^N J_{ij}(a,a_j) \right), 
\end{equation}
with $N$ the length of the sequence, $h_i$ the fields acting on site $i$ and $J_{ij}$ the couplings between site $i$ and $j$.
Since the model is trained using pseudo-likelihood maximization over a dataset of protein sequences, we directly use the \textit{plmDCA} package \cite{Ekeberg2013,Ekeberg2014} to infer the parameters $h_i$ and $J_{ij}$. 
Usually, the couplings $\textbf{J}$ which maximize the likelihood of training data are then symmetrized. 
As in the previous sections, we skip this step and use asymmetric couplings.

Before proceeding, it is worth highlighting some key differences between the approach used in this specific case, based on the \textit{plmDCA} package, and the one adopted throughout the rest of the section:
1) In proteins the alphabet is composed by 21 letters (20 amino acids plus the gap symbol '-') instead of just $\{-1,+1\}$. 
2) As it is clear from Eq.~\ref{eq:plmDCA}, in this case we have not only couplings between variables, but also fields acting on each single site.
3) A regularization on the fields and on the couplings is added, which bounds the values of $\textbf{h}$ and $\textbf{J}$.
4) The dynamics is sequential instead of parallel.

As we will see, the qualitative results are nonetheless comparable.
We focus in particular on two protein families, the DNA-Binding Domain (DBD) and the Beta Lactamase, with length $N$ of 76 and 202 amino acids, respectively.
For each protein family, we infer the parameters $\textbf{h}$ and $\textbf{J}$ following the procedure described in \cite{Ekeberg2013,Ekeberg2014} and we study the behavior of the system under a zero-temperature dynamics.
The evolution starts from a certain sequence $\textbf{a}^{\rm ini}$ of amino acids and proceeds according to Eq.~\ref{eq:dyn_p_zero_T}.
As the dynamics takes place at $T=0$, the system eventually reaches a stable configuration and the evolution halts, yielding a final sequence $\textbf{a}^{\rm fin}$.
One can then compute the overlap $q(\textbf{a}^{\rm ini}, \textbf{a}^{\rm fin}) = 1/N \sum_{n=1}^N \delta_{a^{\rm ini}_n,a^{\rm fin}_n}$ between the final state and the initial one, with $\delta_{a,b}$ the Kronecker delta function.

This procedure is repeated for different values of $P$, the number of sequences used in the training of the model.
For each value of $P$, we choose 2000 starting sequences drawn uniformly (with repetition) from the training set and we let them evolve, comparing the resulting stable states with the starting condition.
We then repeat the same steps drawing this time the initial sequences from the test set (the unseen examples).

The results are plotted in Fig.~\ref{fig:Proteins}. 
Similarly to what observed in the other cases, we see that, for small values of $\alpha$, under the zero-temperature dynamics the model does not move from its initial state when starting from a sequence in the training set. 
This memorization region holds up to $\alpha$ between 1 and 10. 
For larger values, the training sequences are not anymore fixed points of the dynamics and the overlap quickly saturates to the large-load value. 
Starting from the test sequences, the overlap follows an opposite trend: for small values of $\alpha$, $m_{\mathrm{F}}$ is close to the average overlap between natural sequences (dashed line), while as the load increases it increases as well. 
For large values of $\alpha$, $m_{\mathrm{F}}$ saturates to the same point as the one obtained starting from the training sequences. 
Differently from the previous examples, with proteins we observe $m_{\mathrm{F}}$ around 0.55 and 0.6 in the generalization phase, meaning that the attractors of the model are correlated with train and test examples even if the model is unable to retrieve them (which was expected, as these datasets should be much harder).
This is clear when comparing the final values of $m_{\textrm{F}}$ with the average pairwise distance between natural sequences (dashed line).

\begin{figure}
\centering
\includegraphics[width=0.5\textwidth]{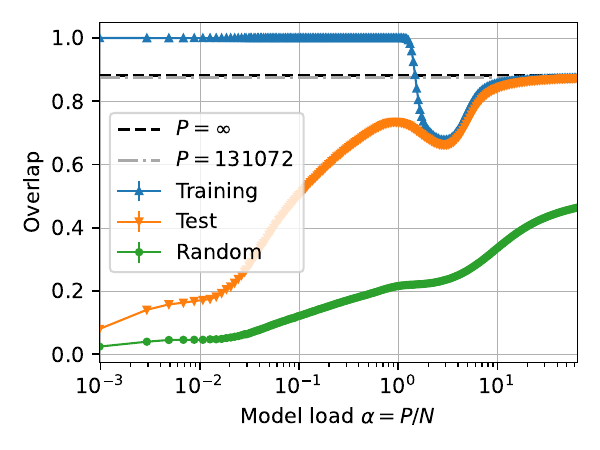}
\caption{\textbf{Generalization transition for configurations of an Edwards-Anderson model.} \textit{Data points:} average final overlap obtained running the zero-temperature dynamics of the pseudo-likelihood model starting from training data (blue), from test data (orange) and random configurations (green) as a function of the number of the load $\alpha = P/N$. \textit{Horizontal lines:} the final average overlaps obtained when using the full dataset $M$ (gray, dotted) and the original couplings of the model (black, dashed). We considered a system of size $N = 32\times 32 = 1024$ spins. Error bars have sizes comparable to data points.}
\label{Fig:ea2d}
\end{figure}

\subsection{Edwards-Anderson model in two dimensions}
We consider data sampled at thermodynamic equilibrium from the Edwards-Anderson (EA) \cite{edwards1975theory} spin glass model in two dimensions (see appendix \ref{sec:appendix_EA} for details on the model, the data generation process and the training dynamics).
From Fig.~\ref{Fig:ea2d} we see that there is a first memorization region (up to $\alpha > 1$) in which the training data are fixed points of the dynamics. At higher values of $\alpha$, one moves to a generalization regime in which one is able to infer with good approximation the couplings \cite{nguyen2017inverse}, so that performing the zero-temperature dynamics on the pseudo-likelihood-inferred couplings is approximately equivalent to using directly the original $J_{ij}$ of the model. This is further illustrated by the gray and black lines, which correspond to the final overlaps obtained using couplings inferred from the $M$ configurations (gray line) and the original couplings of the model (black lines). As a consequence of this fact, the norm of the couplings does not diverge during training in the $\alpha \to \infty$ limit, and the same is expected to hold at $\alpha < \infty$. Nonetheless, the behavior of the systems as an AM is similar to that of the previous sections and to the MNIST case, which we discuss below.

\section{Discussion}
\label{sec:conclusions}

To study the transition from overfitting to generalization in energy-based probabilistic models, we trained a model by maximizing the pseudo-likelihood approximation and run a recurrent parallel dynamics at zero temperature. We found that, in this setting, the model behaves as an AM regardless of the symmetry of the coupling. Furthermore, we found that this AM present attractor that are close to previously unseen test examples, and we propose this phenomenon as a new paradigm for generalization.

\paragraph*{Relation to maximum likelihood}
This work raises the question of whether AM appear in inverse problems only due to pseudo-likelihood, or if instead it is a more general phenomenon in energy-based models. 
To the best of our knowledge, the minima produced by maximizing the likelihood  in energy-based models have no clear theoretical connection with training examples. Notably, in the few situations where theory is available, attractors appear uncorrelated with training examples
\cite{decelle2021restricted,catania2025theoretical}, but the propensity of generating samples near training examples is well-known in practice \cite{bereux2025fast, bonnaire2025diffusion}. In light of this apparent contradiction, extending our findings to maximum likelihood training is particularly important.  

\paragraph*{Finite temperature and hidden units}
To better answer the question of overfitting to generalization in energy-based models, two extensions of this work are desirable. First, to repeat the study at finite temperature and understand if the structure of basins of attraction that we found at zero temperature is still relevant to describe a sampling process. 
Second, it would be interesting to include hidden variables, to better describe common architectures such as Restricted Boltzmann Machines, but a preliminary study would be required since it is unclear if the generalization transition identified in \cite{kalaj2024random} would still happen in the same way when hidden variables are present.

\paragraph*{Applications to deep learning}
Modern self-supervised problems like generative diffusion and self-attention mechanism that have been directly related to AMs \cite{ambrogioni2024search} and pseudo-likelihood training \cite{rende2024mapping,d2024self}. Our work contributes to this perspective by showing that even the simplest possible architecture trained in self-supervised fashion becomes an AM. Indeed, the perspective that we adopted on generalization is more natural in the setting of self-supervised learning, where one measures generalization as the number of variables correctly predicted given the others on a test set. Our work adds new pieces to understand quantitatively the transition to generalization, and in particular our use of the $P(\Delta)$ may be extended to more complicated architectures. 

\paragraph*{Applications to neuroscience}
In section \ref{sec:PLisAM} we argued on why pseudo-likelihood produces and AM using the distribution of stabilities $P(\Delta)$ with random memories. This argument also showed that, at early training the model implements the Hebbian learning, and as training goes on, the capacity of the AM increases. 
Given that Hebbian learning has been considered as a model for biological neurons learning in response to external stimuli for a long time \cite{hebb1949organization}, we remark that the maximization of pseudo-likelihood retains some elements of biological plausibility. First, the natural asymmetry of the coupling matrix. Second, the factorization of the loss makes each neuron optimize its own local loss, without operations that involve other neurons. 
Due to these properties, it would be worth to study whether the pseudo-likelihood maximization offers a model compatible with the synaptic plasticity of a network of biological neurons.

\section{Acknowledgments}

We thank Stefano Crotti and Enrico Ventura for pointing out relevant literature. 
We also thank Federico Ricci-Tersenghi for providing the code for the parallel tempering simulations. LMDB acknowledges the support of ``Bando Ricerca Scientifica 2024 - Avvio alla Ricerca" of Sapienza University.
SR acknowledges the support of FIS (Italian Science Fund) 2021 funding scheme (FIS783 - SMaC - Statistical Mechanics and Complexity) from MUR, Italian Ministry of University and Research and from the PRIN funding scheme (2022LMHTET - Complexity, disorder and fluctuations: spin glass physics and beyond) from MUR, Italian Ministry of University and Research.
MN acknowledges the support of PNRR MUR project PE0000013-FAIR.

\bibliography{references}
\bibliographystyle{unsrt}

\appendix

\onecolumngrid

\newpage
\section*{Appendix}

\section{Computation of the stabilities}
\label{sec:replicaGardner}

Given the loss function
\[
L(\pmb{w}) \;=\; \sum_{\mu=1}^{
\alpha N} V(\Delta^\mu),
\qquad
\Delta^\mu \;\equiv\;
y^{\mu}\!\left(
      \pmb{w}\!\cdot\!\pmb{x}^{\mu}
\right),
\]
where $\Delta^\mu$ is the \emph{stability} of pattern~$\mu$, we wish to determine the margin distribution
$P_{\alpha, \beta}(\Delta)$ averaged over the dataset
$\{\pmb{x}^\mu\}_{\mu=1}^{\alpha N}, \; x^\mu_i = \pm 1 \; \text{randomly}$, and over weight vectors $\pmb{w} \in \mathbb{R}^N$ drawn from the Boltzmann measure on the sphere
$\lVert\pmb{w}\rVert=1$, with density proportional to $\exp[-\beta L(\pmb{w})]$
The integration element
\[
d\mu(\pmb{w}) \;=\;
\frac{d^{N}\pmb{w}}{\Omega_{N}}\,
\delta\!\bigl(\lVert\pmb{w}\rVert^{2}-1\bigr),
\qquad
\Omega_{N}=\frac{2\pi^{N/2}}{\Gamma(N/2)},
\]
is the \emph{uniform measure} on that sphere. 
Explicitly, we write the stability distribution as
\begin{equation}
\label{eq:Pbeta_def}
P_{\alpha,\beta}(\Delta) \;=\;
\left\langle
\frac{
      \displaystyle
      \int d\mu(\pmb{w})\,
      e^{-\beta \sum_{\mu=1}^{\alpha N}
          V\!\bigl(y^\mu\pmb{w}\cdot\pmb{x}^\mu\bigr)}
      \,\delta\!\Bigl(
          \tfrac{y^{1}}{\sqrt{N}}\pmb{w}\cdot\pmb{x}^{1}-\Delta
      \Bigr)}
     {
      \displaystyle
      \int d\mu(\pmb{w})\,
      e^{-\beta \sum_{\mu=1}^{\alpha N}
          V\!\bigl(y^\mu\pmb{w}\cdot\pmb{x}^\mu\bigr)}
      }
\right\rangle_{\!\pmb{x}^\mu},
\end{equation}
where, by symmetry, we insert the $\delta$–function for the first pattern only.

Applying the replica trick, we (non-rigorously) rewrite \eqref{eq:Pbeta_def} as
\begin{equation}
\label{eq:margin_dist}
P_{\alpha,\beta}(\Delta) \;=\;
\lim_{n\to 0}\Bigl\langle
 \int\!\prod_{a=1}^{n} d\mu(\pmb{w}^{a})\,
      e^{-\beta\sum_{\mu,a}
      V\!\bigl(y^\mu\pmb{w}^{a}\!\cdot\!\pmb{x}^\mu\bigr)}
      \,\delta\!\Bigl(
          y^{1}\pmb{w}^{1}\!\cdot\!\pmb{x}^{1}-\Delta
      \Bigr)
\Bigr\rangle_{\!\pmb{x}^\mu}\!.
\end{equation}

Focusing on minimisers of the loss ($\beta\to\infty$), by assuming that $V(\Delta)$ has a unique minimum in $\Delta$ and assuming symmetry between replicas \cite{mezard1987spin}, we obtain
\[
P_{\alpha,\beta\to\infty}(\Delta)
 \;=\;
 \int_{-\infty}^{\infty}\!\frac{dt}{\sqrt{2\pi}}
      e^{-t^{2}/2}\,
      \delta\!\Bigl(\Delta-\Delta_{0}\bigl(t,\bar{x}(\alpha)\bigr)\Bigr),
\]
with
\[
\Delta_{0}(t,x)\;=\;
\mathop{\arg\min}_{\Delta}
      \Bigl[V(\Delta)+\tfrac{(\Delta-t)^{2}}{2x}\Bigr],
\qquad
1 \;=\; \alpha
      \int \mathcal{D}y\,
      \bigl[\Delta_{0}(y,\bar{x}(\alpha))-y\bigr]^{2},
\]
and $\mathcal{D}y = e^{-y^{2}/2}dy/\sqrt{2\pi}$.

\section{More examples of retrieval in MNIST}

\begin{figure*}[h]
\begin{center}

\includegraphics[width=0.6\textwidth]{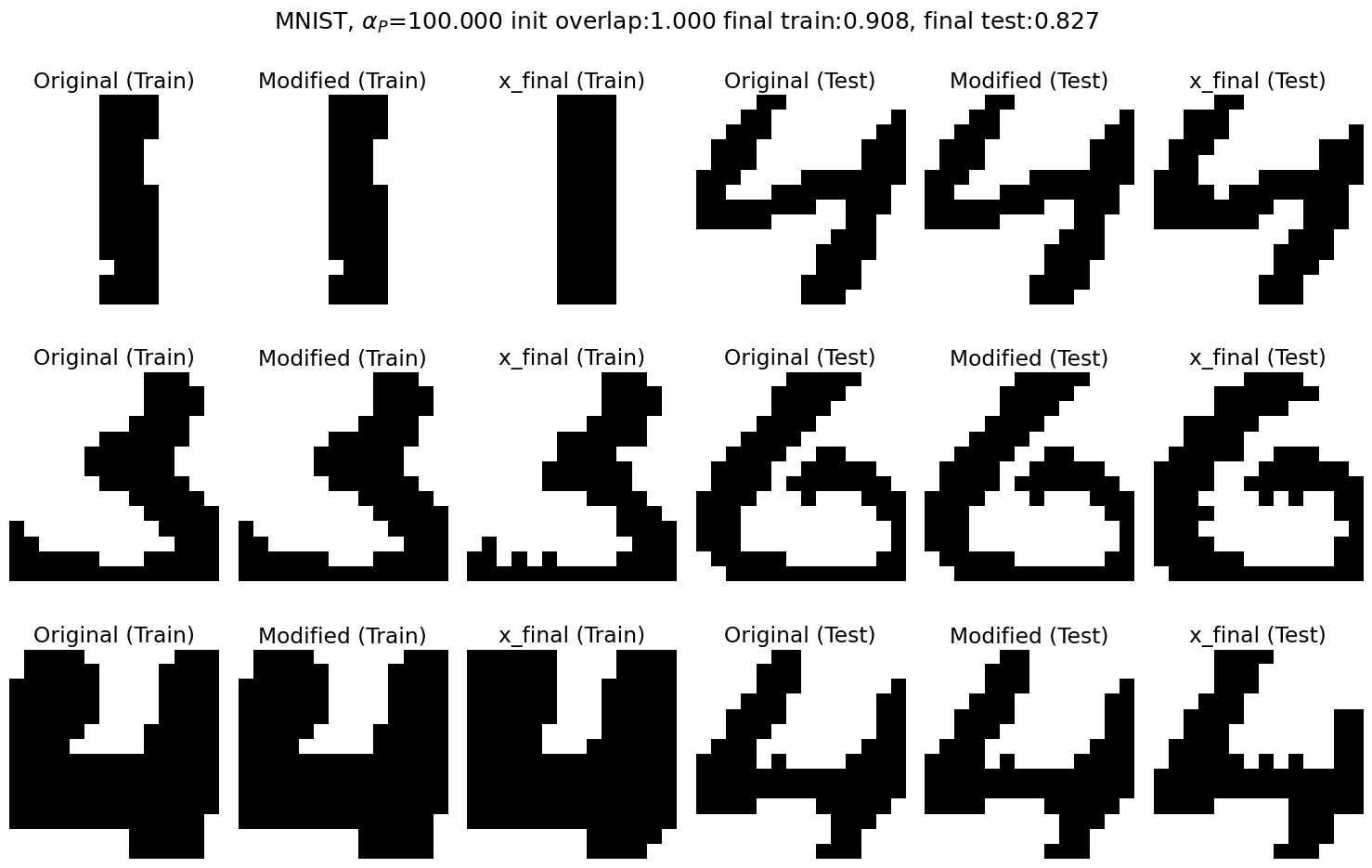}
\includegraphics[width=0.6\textwidth]{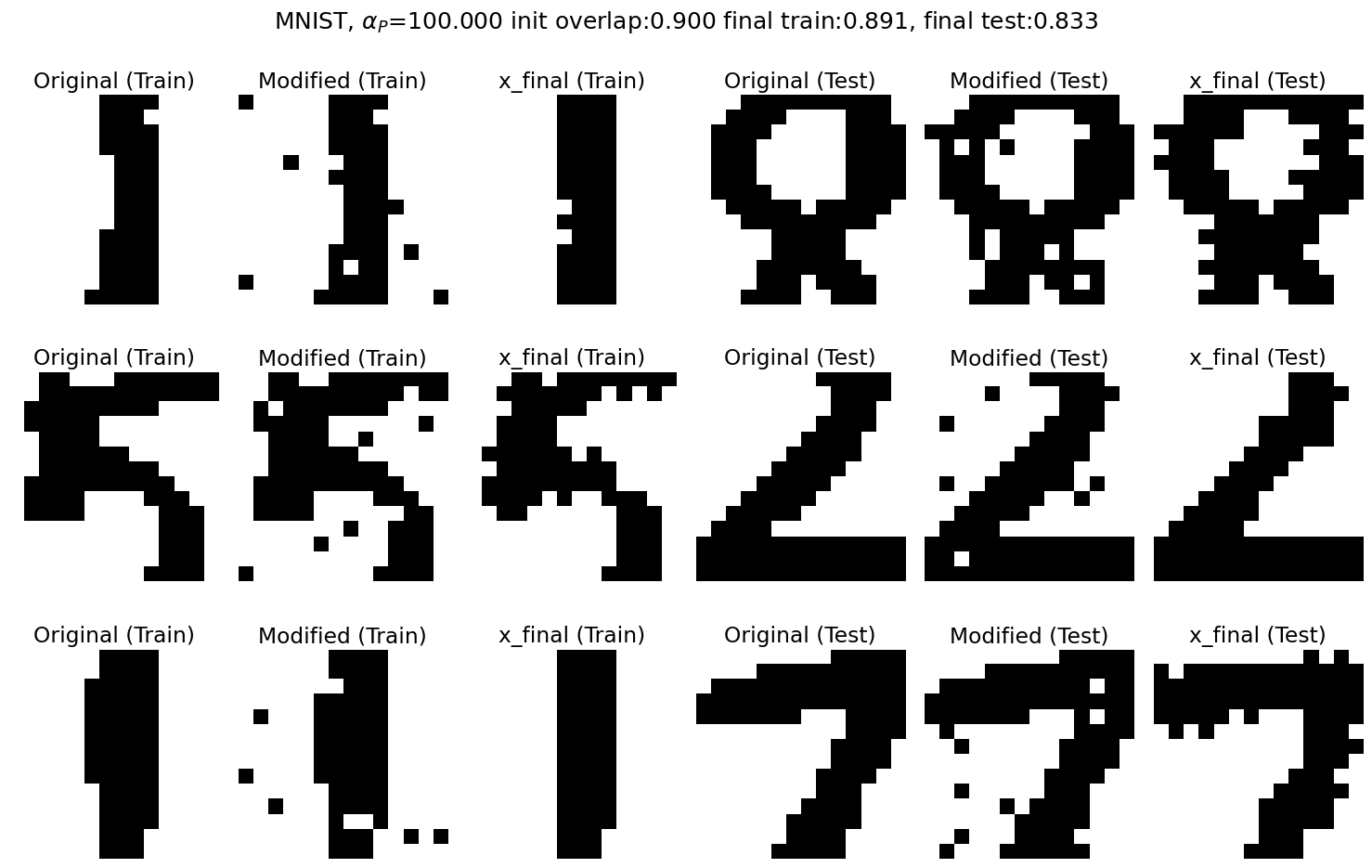}
\includegraphics[width=0.6\textwidth]{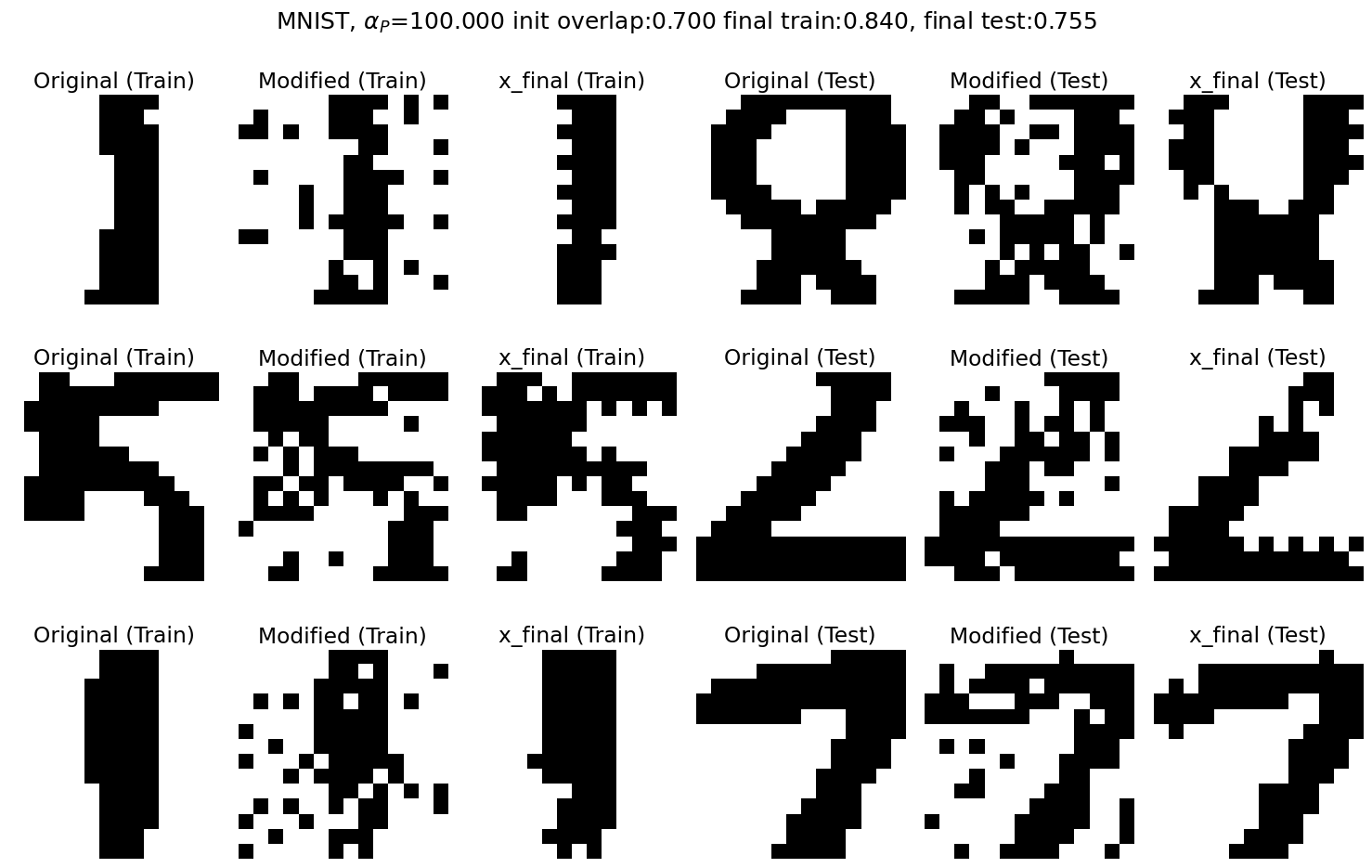}

\end{center}
\caption{Examples of retrieval of test and train MNIST images with various initial overlaps.}
\label{fig:mnsit_examples}
\end{figure*}

In figure \ref{fig:mnsit_examples} we provide more examples of attractors starting from train and test examples. In particular, we visualize how stable these attractors are with respect to noisy version of the examples.

\label{sec:appendix_mnist}

\section{Details of the Edwards-Anderson model}
\label{sec:appendix_EA}

In the EA model, there is a set of $N$ interacting Ising variables $\bs = \{\s_1,\dots,\s_N\}$, $\s_i = \pm 1$, $i = 1, \dots, N$,  distributed according to the Gibbs-Boltzmann distribution
\begin{equation}\label{eq:GB_distribution}
    P_\text{GB}(\bs) = \frac{e^{-\beta \mathcal{H}(\bs)}}{\mathcal{Z}(\beta)} \ ,
\end{equation}
where $\beta = 1/T$ is the inverse of the temperature $T$ and the (temperature-dependent) normalization constant $\mathcal{Z}(\beta)$ is the partition function. The Hamiltonian $\mathcal{H}$ is taken as \begin{equation}\label{eq:spin_glass_H}
    \mathcal{H} = - \sum_{i} H_i \sigma_i 
    -\sum_{ (i,j) \in \mathcal{E}} J_{ij} \sigma_i \sigma_j,
\end{equation}
and is parametrized by as set of external local fields $H_i$ and pairwise couplings $J_{ij}$.
These quenched parameters are chosen as i.i.d. random variables, e.g. Gaussianly distributed
$H_i \sim \mathcal{N}(0,H^2)$ and
$J_{ij} \sim \mathcal{N}(0,J^2)$. Here, $\mathcal{E}$ is the edge set of the graph representing the topology on which the model is defined. In our case, it is a square lattice of side $L$ (so that $N = L^2$) with open boundary conditions.

We considered a system with $L = 32$, $T = 0.99$, zero external fields ($H_i = 0 \; \; \forall i$) and unitary coupling variance ($J = 1$). We generated a set of $M = 2^{17} = 131072$ total configurations distributed according to \eqref{eq:GB_distribution} using the standard Parallel Tempering algorithm \cite{hukushima1996exchange}.

For $P$ ranging from 1 to $M/2 = 65536$, we then proceeded to extract $P$ configurations at random from the $M$ total configurations. For each set of $P$ configurations, we trained a pseudo-likelihood model using a gradient descent with learning rate 100 and exponential decay (the decay factor was set to 0.999 and it only slightly reduces the learning during training), trained for 1000 epochs. We then performed a zero-temperature dynamics starting from the training data, from other $P$ configurations chosen at random (\textit{test data}) and from configurations generated uniformly at random from the $N$-dimensional hypercube. We repeated the whole process 100 times and averaged the results. The final mean overlaps reached after 1000 zero-temperature steps are plotted as a function of $\alpha = P/N$ in Fig. \ref{Fig:ea2d}.

\section{Details specific to the datasets of protein sequences}
\label{sec:appendix_proteins}
We test the procedure described for the other datasets also on data coming from protein sequences. 
Notice that, in this case, we are not generating synthetic data anymore, instead, we are taking as train and test sets some sequences of amino acids that are found in nature.  
From each protein family we produce a Multiple Sequence Alignment (MSA) in order to have all the sequences with the same length $N$ and aligned between them. 
The MSA for the two families are obtained in the following way:
\begin{itemize}
    \item
    For DNA-Binding Domain (DBD) we used as seed the alignment from~\cite{park2022epistatic} (221 sequences) and ran the HMMER~\cite{eddy2009new} command {\tt hmmsearch} on the {\tt uniref90} database~\cite{uniref2007}, excluding sequences with more than 20\% gaps.
    \item For Beta-Lactamase we used the same alignment as in~\cite{bisardi2022modeling}.
\end{itemize}
In order to account for the bias due to the presence of many similar sequences in the natural MSA, we assign a weight $w_s = 1/n_s$ to each sequence $s$, with $n_s$ the number of different sequences in the alignment closer than 80\% in length to $s$. 
The value $M_{\rm eff} = \sum_s w_s$ is the effective number of sequences in the alignment.
This is a standard procedure when dealing with this kind of data~\cite{Ekeberg2013,Ekeberg2014}.
For the two protein families under consideration here, we have length $N=76$, number of sequences $M=13310$ ($M_{\rm eff}=3153$) for the DNA-Binding Domain (DBD) and $N=202$, $M=18334$ ($M_{\rm eff}=6875$) for the Beta-Lactamase.
From this MSA we select a set of $P$ sequences that we use to train the model, drawn from the full MSA with a probability proportional to their weight.
We choose $P$ between 1 and $M_{\rm eff}$ for each family.

\end{document}